\documentclass[12pt]{aastex}
\usepackage{emulateapj5}

\newcommand\kpc{\:\rm{kpc}}

\newcommand\VEL{\:{\rm km\:s^{-1}}}

\newcommand\OIGS{\:{\rm ergs\:cm^{-2}\:s^{-1}\:\AA^{-1}}}

\newcommand{\kms}{km~s$^{-1}$}

\def\deg{\hbox{$^\circ$}}

\begin{document}

\title{Overview of the \emph{Far Ultraviolet Spectroscopic Explorer} Mission}

\author{H.W.~Moos\altaffilmark{1},
W.C.~Cash\altaffilmark{2},
L.L.~Cowie\altaffilmark{3},
A.F.~ Davidsen\altaffilmark{1},
A.K.~Dupree\altaffilmark{4},	
P.D.~ Feldman\altaffilmark{1},
S.D.~Friedman\altaffilmark{1},
J.C.~Green\altaffilmark{2},
R.F.~Green\altaffilmark{5},
C.~Gry\altaffilmark{6,20}, 
J.B.~Hutchings\altaffilmark{7},
E.B.~Jenkins\altaffilmark{8}, 
J.L.~Linsky\altaffilmark{21},
R.F.~Malina\altaffilmark{6,9},
A.G.~Michalitsianos\altaffilmark{10,13},
B.D.~Savage\altaffilmark{11}, 
J.M.~Shull\altaffilmark{2,21},
O.H.W.~Siegmund\altaffilmark{12},
T.P.~Snow\altaffilmark{2},
G.~Sonneborn\altaffilmark{13},	
A.~Vidal-Madjar\altaffilmark{14},
A.J.~Willis\altaffilmark{15},
B.E.~Woodgate\altaffilmark{13},
and D.G.~York\altaffilmark{16}}
\affil{\emph{FUSE} Science Team}

\author{T.B.~Ake\altaffilmark{1},	
B-G~Andersson\altaffilmark{1},
J.P.~Andrews\altaffilmark{2},
R.H.~Barkhouser\altaffilmark{1},
L.~Bianchi\altaffilmark{1,22},
W.P.~Blair\altaffilmark{1},
K.R.~Brownsberger\altaffilmark{2},
A.N.~Cha\altaffilmark{1},	
P.~Chayer\altaffilmark{1,17}, 
S. J. Conard\altaffilmark{1},
%L.J.~Frank\altaffilmark{24}
A.W.~Fullerton\altaffilmark{1,17},
G.A.~Gaines\altaffilmark{12},
R.~Grange\altaffilmark{6},
M.A.~Gummin\altaffilmark{12},
G.~Hebrard\altaffilmark{14},
G.A.~Kriss\altaffilmark{1,18},
J.W.~Kruk\altaffilmark{1},
D.~Mark\altaffilmark{19},
D.K.~McCarthy\altaffilmark{1,23},
C.L.~Morbey\altaffilmark{7},
R.~Murowinski\altaffilmark{7},
E.M.~Murphy\altaffilmark{1},
W.R.~Oegerle\altaffilmark{1},
R.G.~Ohl\altaffilmark{1},
C.~Oliveira\altaffilmark{1},
S.N.~Osterman\altaffilmark{2},
%K.C.~Roth\altaffilmark{1},
D.J.~Sahnow\altaffilmark{1},
M.~Saisse\altaffilmark{6},
K.R.~Sembach\altaffilmark{1},
H.A.~Weaver\altaffilmark{1},
B.Y.~Welsh\altaffilmark{12},
E.~Wilkinson\altaffilmark{2},
and W.~Zheng\altaffilmark{1}}
\affil{\emph{FUSE} Instrument Team and \emph{FUSE} Science Operations Team}

\altaffiltext{1}{Department of Physics \& Astronomy, The Johns 
Hopkins University, Baltimore, MD  21218} 
\altaffiltext{2}{Center for Astrophysics and Space Astronomy, Department of 
Astrophysical and Planetary Sciences, University of Colorado, CB 389, Boulder, 
CO 80309}
\altaffiltext{3}{Institute for Astronomy, University of Hawaii, 2680 Woodlawn 
Drive, Honolulu, HI 96822}
\altaffiltext{4}{Center for Astrophysics, 60 Garden Street, Cambridge, MA 
02138}
\altaffiltext{5}{National Optical Astronomy Observatories, Tucson, AZ 85726}
\altaffiltext{6}{Laboratoire d'Astronomie Spatiale, B.P.8, 13376 Marseille cedex 
12, France.}
\altaffiltext{7}{Herzberg Institute of Astrophysics, National Research 
Council of Canada, 5071 West Saanich Road, Victoria, BC V8X 4M6, Canada}
\altaffiltext{8}{Princeton University Observatory, Princeton, NJ 08544}
\altaffiltext{9}{Center for EUV Astrophysics, University of California at 
Berkeley, 2150 Kittredge Street, Berkeley, CA 94720-5030}
\altaffiltext{10}{Deceased}
\altaffiltext{11}{Department of Astronomy, University of Wisconsin, Madison, 
WI 53706}
\altaffiltext{12}{Space Sciences Laboratory, University of California, Berkeley, 
CA 94720-7450}
\altaffiltext{13}{Laboratory for Astronomy and Solar Physics, NASA/GSFC, Code 
681, Greenbelt, MD 20771}
\altaffiltext{14}{Institut d'Astrophysique de Paris, CNRS, 98 bis 
bld Arago, F-75014 Paris, France}
\altaffiltext{15}{Department of Physics and Astronomy, University College 
London, Gower St., London, UK}
\altaffiltext{16}{University of Chicago, Department of Astronomy and 
Astrophysics, 5640 South Ellis Avenue, Chicago, IL 60637}
\altaffiltext{17}{Primary affiliation:  Dept. of Physics \& Astronomy, 
University of Victoria, P.O. Box 3055, Victoria, BC, V8W 3P6, Canada.} 
\altaffiltext{18}{Space Telescope Science Institute, Baltimore, MD 21218}
\altaffiltext{19}{Swales Associates, 5050 Powder Mill Road, 
Beltsville, MD 20705}
\altaffiltext{20}{ISO Data Center, ESA Astrophysics Division, PO Box 50727, 
28080 Madrid, Spain}
\altaffiltext{21}{JILA, University of Colorado and NIST Campus Box 440, Boulder, 
CO 80309-0440}
\altaffiltext{22}{Astronomical Observatory of Torino, I-10025 Pino Torinese 
(TO), Italy}
\altaffiltext{23}{Present address:  Swales Associates, 5050 Powder Mill Road, 
Beltsville, MD 20705}

\begin{abstract}

The \emph{Far Ultraviolet Spectroscopic Explorer} satellite observes light in 
the far-ultraviolet spectral region, 905 -- 1187 \AA\ with high spectral 
resolution. The instrument consists of four coaligned prime-focus 
telescopes and Rowland spectrographs with microchannel plate detectors.  Two of 
the telescope channels  use Al:LiF coatings for optimum reflectivity from 
approximately 1000 to 1187 \AA\ and the other two use SiC coatings for optimized 
throughput between 905 and 1105 \AA.  The gratings are holographically ruled to 
largely correct for astigmatism and to minimize scattered light.  The 
microchannel plate detectors have KBr photocathodes and use photon counting to 
achieve good quantum efficiency with low background signal.  The sensitivity is 
sufficient to examine reddened lines of sight within the Milky Way as well as 
active galactic nuclei and QSOs for absorption line studies of both Milky Way 
and extra-galactic gas clouds.  This spectral region contains a number of key 
scientific diagnostics, including \ion{O}{6}, \ion{H}{1}, \ion{D}{1} and the 
strong electronic transitions of H$_2$ and HD.

\end{abstract}


\section{Introduction}

The \emph{Far Ultraviolet Spectroscopic Explorer} (\emph{FUSE}) is a NASA 
astronomy mission, developed in cooperation with the Canadian Space Agency and 
the Centre National d'Etudes Spatiales of France, that is exploring the 
far-ultraviolet (FUV) universe from 905 to 1187 \AA\ with high spectral 
resolution.  \emph{FUSE} was launched 1999 June 24 on a Delta II rocket.  Early 
Release Observations, which are the basis of the papers in this issue of the 
Astrophysical Journal Letters, began in 1999 October and regular science 
operations commenced in 1999 December.  
%\begin{figure*}
%\epsscale{1.5} 
%\plotone{fig1.eps}
%\caption{\emph{FUSE} spectrum of the CSPN K1-16, 912--992 \AA.  Data are 
%from the SiC channel on the detector 1 side.  Transitions in the intervening 
%gas are marked below the spectrum.  Transitions intrinsic to the central star 
%are marked above the spectrum.}
%\end{figure*}

After the decommissioning of the \emph{Copernicus} Mission \citep{spi75} in 
1981, it was clear that a follow-on mission with much higher sensitivity and a 
velocity resolution comparable to or better than that of \emph{Copernicus} 
($\sim15$ \kms\ FWHM
\footnote{Full-width-half-maximum of a spectral profile.  All resolution widths 
 are FWHM unless specifically stated otherwise.})
 was highly desirable.  The utilization of modern detectors and mirror 
technology opens the spectral bandpass from the short-wavelength cutoff of the 
\emph{Hubble Space Telescope} (\emph{HST}) down to the \ion{H}{1} 
photoionization limit at 912 \AA\ for observations at distances far beyond the 
$\sim1\kpc$ limit for routine \emph{Copernicus} measurements.  \emph{FUSE} 
received strong support from two decadal survey committees \citep{fie82,bah91}.  
Although cost considerations restricted the telescope for such a mission to 
roughly the one meter class, modern detectors that simultaneously cover most of 
the bandpass made high spectral resolution measurements with excellent 
sensitivity possible. In addition, by using modern mirror technology, the 
bandpass could be extended below $\sim1000$ \AA\ where the \emph{Copernicus} 
sensitivity was limited by the drop in reflectivity of mirrors overcoated with 
Al and a thin covering of LiF (Al:LiF).  Finally, the installation of the Space 
Telescope Imaging Spectrograph (STIS) \citep{woo98} on the \emph{HST} with a 
lower limit of $\sim1150$ \AA\ meant that the FUSE mission could be designed 
primarily for the wavelengths below the STIS limit.

Exploration of this spectral region since the \emph{Copernicus} 
mission has been limited.  The Voyager ultraviolet spectrometers \citep{bro77} 
obtained low resolution ($\Delta\lambda\ge18$ \AA) spectra of a number of 
sources, e.g. \citep{hol91}. Several missions covering this 
wavelength range have been flown as space shuttle sortie missions with 
operational lifetimes of up to two weeks.  These included the Hopkins 
Ultraviolet Telescope (HUT) \citep{dav92, kru95}, the Orbiting and Retrievable 
Extreme Ultraviolet Spectrometers (ORFEUS) \citep{gre91,hur98}  and the 
Interstellar Medium Absorption Profile Spectrograph (IMAPS) \citep{jen96}.  
These missions performed a wide range of important studies.  However, they were 
precluded from carrying out more detailed and comprehensive studies by the 
limited length of the missions as well as limits to the velocity resolution or 
sensitivity. 

This article presents an overview of the \emph{FUSE} mission.   A discussion of 
the scientific background is followed by a description of the mission. The 
following article \citep{sah00} discusses the on-orbit performance of the 
\emph{FUSE} satellite in more detail.

\section{Scientific Background}

The density of strong spectral absorption lines per wavelength interval rises 
sharply in the FUV because a large number of astrophysically 
important species have strong transitions in this region.  This is not suprising 
for neutral species, since the photon energy at these wavelengths approaches one 
rydberg (13.6 eV).  For some species, the access is unique.  These include all 
of the resonance lines of H and D except for Ly$\alpha$,  the strong resonance 
lines of \ion{C}{3}, \ion{O}{6} and \ion{S}{6}, as well as the electronic 
ground-state absorption bands of H$_{2}$ and HD, including the Lyman and the 
Werner bands.  With modest redshifts, extreme ultraviolet (EUV) transitions 
become observable including \ion{O}{5} $\lambda$629.73, 
\ion{Ne}{8} $\lambda\lambda$770.4, 780.32 and \ion{Mg}{10} 
$\lambda\lambda$609.85, 625.28, which provide access to the hot intergalactic 
medium \citep{ver94}.

As an example of the rich variety of spectral features found in the \emph{FUSE} 
bandpass, Figure 1 displays a  spectrum of the central star of the planetary 
nebula (CSPN) K1-16 (V$=$15) covering 912--992 \AA\ obtained while  mapping the 
large 
spectrograph entrance aperture during on-orbit checkout.  The spectra were 
corrected for the wavelength shifts induced by the target motion in the aperture 
and coadded to give a total integration time of 27 ks.  A large number of 
transitions are apparent within this relatively narrow spectral region. 
Identified transitions due to the intervening gas (including nebular gas) are 
marked below the spectrum and include those of H$_{2}$, \ion{H}{1}, \ion{O}{1}, 
\ion{N}{1}, \ion{Si}{2}, \ion{P}{2},  \ion{C}{3}\ and \ion{S}{6}.  Above the 
spectrum, stellar lines due to transitions between excited states of 
\ion{He}{2}, \ion{C}{4} and \ion{O}{6} are marked with identifications based on 
those in representative PG 1159 stars \citep{kru98}.  
  
A number of the papers in this issue demonstrate the unique capabilities of the 
\emph{FUSE} mission for studies of the  interstellar and intergalactic gas.  For 
example, the \ion{O}{6} $\lambda\lambda$1032, 1038 transitions are extremely 
sensitive tracers of hot gas in the Galactic halo shock-heated  by supernova 
explosions in the disk \citep{sav00}.  Several other papers 
\citep{mur00,oeb00,sem00} use these transitions to study the properties of hot 
gas both in and outside of the Milky Way.  At the other temperature extreme, the 
strong Lyman and Werner transitions provide exceptional sensitivity to cold 
molecular hydrogen \citep{shu00a,sno00, fer00}.  Additional interstellar medium 
studies include the ionization balance of the local interstellar medium 
\citep{jen00} and a line of sight through the Milky Way to the LMC 
\citep{fri00}.  Studies of the abundance ratios of deuterium to hydrogen are a  
major objective of the \emph{FUSE} mission \citep{moo98}.  These 
studies will provide a much better understanding of the astration cycle 
whereby deuterium is destroyed by stellar processes concurrent with metal 
production.  Lines of sight providing a wide variety of astrophysical 
environments are planned, including the local interstellar medium, the Milky Way 
disk and halo, high velocity clouds, and the local intergalactic medium.  
Observations for the deuterium program began in the early spring 2000 and 
deuterium measurements will be reported at a later date.

Studies of winds from massive stars in the LMC and the SMC 
\citep{bia00,cro00,ful00,mas00} use the \ion{O}{6}\ transitions to probe the 
highest temperature gas and show that it exists throughout the wind.  These 
studies confirm significant differences between the properties of the winds 
of otherwise similar stars in LMC and SMC, which may indicate significant 
differences in stellar evolution and fueling of the interstellar medium.  
Even at this early stage of the mission, \emph{FUSE} has been used for a wide 
variety of studies.  Note the articles on the active cool star AB Dor 
\citep{ake00},  the discovery of P and Fe in a hot white dwarf  \citep{cha00}, 
the SNR N49 in the LMC \citep{bla00}, intergalactic Ly$\beta$ absorbers 
\citep{shu00b}, the lack of molecular H$_2$ in I Zw 18 \citep{vid00}, and the 
Seyfert 1 galaxy Mkr 509 \citep{kri00}.  The articles in this issue provide a 
snapshot of the \emph{FUSE} mission a half-year after launch and show the 
potential for future studies.  It is expected that both the Principal 
Investigator Team and Guest Investigators will use the satellite for an even 
wider variety of investigations.

\section{Mission Overview}

The \emph{FUSE} satellite, with a length of $\sim5.5$ m and a mass of 
$\sim1300$ kg, was launched from Cape Canaveral Air Force Station into a 
768 km near-circular orbit with 25\deg\ inclination to the equator.  The 
\emph{FUSE} satellite consists of a single instrument and the spacecraft bus  
built by Orbital Sciences Corporation.  Located below the instrument section, it 
provides power, attitude control, and communications. The primary ground station 
in Puerto Rico and a secondary station in Hawaii provide about eight contacts of 
approximately 10 minutes duration each day. Thus, real time operations are 
limited. All scientific observations are pre-planned in detail and performed 
under autonomous satellite control. The resulting data sets are then stored on 
the spacecraft recorder until contact is made with a ground station.  The ground 
contacts are adequate for transmitting the data to the ground during nominal 
operations, although high data rates require special planning to prevent 
recorder overflow.  A Fine Error Sensor (FES) camera with a 20\arcmin\ field of 
view views one of the telescope Focal Plane Aperture assemblies (FPA).  FES 
images are used for autonomous target field identification.  When the error 
signals from the FES are combined  with those from the spacecraft inertial 
reference unit, pointing with a stability of 0.3\arcsec\ is routinely achieved.  
The positional uncertainty is somewhat larger due to thermal effects on the 
structure \citep{sah00} and uncertainties between target coordinates and 
positions of guide stars selected from the \emph{HST} Guide Star Catalog 
\citep{las90}.

%\begin{figure*}
%\epsscale{0.5} 
%\plotone{fig2.eps}
%\caption{Schematic of the \emph{FUSE} instrument optical system. The telescope 
%focal lengths are 2245 mm and the Rowland Circle diameters are 1652 mm.}
%\end{figure*}

Instrument designs for this spectral region are challenged by the low 
reflectivity of optical coatings:  the number of reflections must be as 
small as possible.  For HUT, this led to an optically economical design:  a 
prime focus telescope and single focusing grating in a Rowland configuration 
\citep{dav92} with a resolution of $\sim300$ \kms.  For high resolution 
spectroscopy, the focal ratio must be $>$ 5 in order to restrict optical 
aberrations.  This leads to very long instruments with extremely large gratings 
that are beyond the scope of most missions.  This was solved for the 
Berkeley spectrograph on the ORFEUS-SPAS missions by sharing the beam 
between three gratings to achieve a velocity resolution of $\sim95$ \kms 
\citep{hur98}. 

For designs with higher resolution, obscuration by the gratings becomes a 
serious consideration.  To solve this, it is necessary to divide the single 
primary mirror into several physically separated optical elements.  Thus, the 
\emph{FUSE} instrument consists of four coaligned prime focus telescopes (off 
axis parabolas, each with 352 mm by 387 mm clear aperture and a focal length of 
2245 mm) feeding light to four Rowland spectrographs \citep{gre94,sah96,fri99}.  
 Figure 2 shows the \emph{FUSE} optical system. In addition to making the 
mission practicable, this approach allowed optimization of the optical coatings 
for different telescope/grating channels.  Two use Al:LiF coatings for optimum 
reflectivity from $\sim1000$ to 1187 \AA\ and the remaining two channels utilize 
SiC coatings for optimized throughput between 905 and 1105 \AA.  

The FPA in each telescope focal plane can be moved in the $z$-direction to 
adjust the spectrograph focus or moved in $x$ (the dispersion direction) to 
coalign the channels.  The $x$ motion can also be used to shift the spectrum 
$\le$ 0.36 \AA\ for the SiC channels and $\le$ 0.39 \AA\ for the LiF channels. 
Each FPA contains three entrance apertures for each spectrograph, corresponding 
to projected angles on the sky of  30\arcsec\ $\times$ 30\arcsec\ (used for most 
observations), 4\arcsec\ $\times$ 20\arcsec\ and 1.25\arcsec\ $\times$ 
20\arcsec.  The entire satellite pointing is changed in order to place 
a target in a particular aperture for all four channels. The grating surfaces 
are spherical with a radius-of-curvature of 1652 mm and are holographically 
ruled at a line density $\sim$ 5767 mm$^{-1}$ for the SiC coated gratings and 
$\sim$ 5350 mm$^{-1}$ for the Al:LiF coated gratings \citep{gre94}.  In addition 
to providing very high ruling density, the holographic ruling process corrects 
for most of the astigmatism \citep{gra92}, decreasing the contribution from the 
detector background signal significantly.  The use of holographic ruling also 
minimizes the  scattered FUV light.  

The requirement for resolution $\le15\VEL$ leads to a dispersion of 1.03 \AA\ 
mm$^{-1}$ for the SiC coated channels and 1.12 \AA\ mm$^{-1}$ for the Al:LiF 
coated channels.  This in turn leads to large-format detectors.  The two 
\emph{FUSE} detectors are multi-segment, two dimensional microchannel plate 
(MCP) detectors with helical double delay line anodes \citep{sig97}.   By 
placing one LiF spectrum and one SiC spectrum in parallel along a single 
detector, we reduced the number of detector systems to two. Each detector 
consists of two segments, each with an active area of 88 $\times$ 10 mm 
curved to approximately match the Rowland Circle; each pair of segments 
is separated by a gap of $\sim$ 10 mm, producing a corresponding wavelength gap. 
The front surface of each MCP is coated with a KBr photocathode to maximize the 
FUV response with a low background count and low sensitivity to out-of-band 
longer wavelength scattered light.  Behind 
each plate are two additional plates that amplify the charge associated 
with each photon event to the equivalent of $\sim 2 \times 10^{7}$ electrons for 
counting and geometric location of each photon event.  The detector pixels 
are $\sim$ 6 $\mu$m in the dispersion direction and 9-16 $\mu$m (depending 
on the detector segment) in the cross dispersion direction, for a full extent of 
roughly 16384 by 1024 pixels. The intrinsic detector resolution, however, is 
determined by the MCP pore size and spacing  (10-15 $\mu$$m$), and by the design 
of the readout electronics; it was shown by ground measurements to be $\sim$20 
$\mu$m in the dispersion direction and $\sim$80 $\mu$m in the cross-dispersion 
direction.

In addition to increasing the effective area, the multiple channel design 
introduces redundancy over most of the spectrum.  The optical design is 
identical for the two Al:LiF coated channels and for the two SiC coated 
channels; the wavelength coverage differs only because the two detectors are 
offset at slightly different locations along the Rowland circles.  SiC1 covers 
905--1090 \AA\ with a gap between the two detector plates at 993--1003 \AA.  
SiC2 covers 917--1104 \AA\ with a gap at 1006--1016 \AA.  LiF1 covers 988--1187 
\AA\ with a gap at 1083--1094 \AA\ and LiF2, 979--1179  \AA\ with a gap at 
1075--1086 \AA. (The transmission of both LiF1 and LiF2 fall sharply below the 
Al:LiF reflectivity cutoff at $\sim 1000$ \AA.)  Thus, the majority of the 
wavelength range 
is covered by two detectors and the central third by all four channels, making 
the system extremely robust against a partial failure of an optical channel or 
even one of the detectors.  In addition, this design aids in distinguishing 
between real features in the data and instrumental artifacts.

The data are recorded in two modes.  For low data rates the $x$-$y$ address of 
each detected photon is listed in order of detection.  A time stamp with an 
accuracy of 8 ms is inserted in the list once a second.  At count rates above 
$\sim 2500$ s$^{-1}$, this mode would require an excessive allocation of 
on-board memory and the spectral image mode is used instead. In this mode, 
images corresponding to the portions of the detectors on which the spectra 
appear are integrated over an exposure.  To further reduce the required memory, 
we bin the data by 8 pixels in the cross-dispersion direction.  In this case, 
there is no photon timing information and on-ground Doppler corrections can only 
be made on the image as a whole;  therefore, each orbit of the observation is 
divided into multiple (typically four) exposures. 

Once the data are transferred from the satellite to the Spacecraft Control 
Center at the Johns Hopkins University, they are passed through level zero 
processing, which sorts the data packets and checks for duplicate or missing 
packets.  The data are then passed to the OPUS pipeline \citep{ros98}, which 
controls the processing of data from ingest until archiving.  OPUS also converts 
the data into FITS format and populates a number of keywords in the header with 
entries from the mission planning database.  OPUS passes the data to the 
calibration pipeline which is a software package specifically designed for 
\emph{FUSE} data. This software takes raw photon address data or spectral image 
data and creates a calibrated, extracted one-dimensional spectrum.  Even though 
the gratings are holographically corrected, a small amount of astigmatism and 
line curvature remain.  This and other effects \citep{sah00} must be removed for 
measurements with the most demanding velocity-resolution requirements.  Although 
not implemented at present, the calibration pipeline has been designed to 
include these corrections in future calibrations of the data.  Finally, 
OPUS collects all the raw and calibrated data files from an observation and 
generates ancillary files needed for ingestion of the data into the 
Multi-Mission Archive at the Space Telescope Science Institute (MAST).  The 
individual steps in the calibration pipeline are described in the \emph{FUSE} 
Data Handbook
\footnote{The \emph{FUSE} Data Handbook \citep{oeg00} is available at 
\url{http://fuse.pha.jhu.edu}} \citep{oeg00}.

\section{Performance}

The on-orbit performance properties of the \emph{FUSE} satellite are discussed 
in detail by \citet{sah00}.  We discuss here the steps taken to maintain an 
effective area sufficient for absorption-line studies of the Milky Way halo and 
extragalactic gas clouds.  We note that the low scattered light leads to 
accurate values of the flux levels in the cores of strongly saturated 
interstellar absorption lines.  Also,  the background signals \citep{sah00} due 
to the  detector dark count  and scattered airglow are sufficiently low that 
very faint objects can be studied to a limited extent.

In the FUV spectral region reflecting optics are very sensitive to contamination 
by organic materials and considerable care in the material selection and 
assembly procedures are necessary.  Also, mirrors and gratings that are 
overcoated with Al and a thin layer of LiF are easily  degraded by exposure to 
air.  The LiF overcoat absorbs water easily and even modest exposures to normal 
laboratory conditions with a relative humidity of 50\%\ will lead to decreases 
in reflectivity \citep{oli99}.  Thus, it was necessary to perform most of the 
instrument assembly and testing with the mirrors and gratings in a dry nitrogen 
environment.  The total exposure to cleanroom air with relative humidity 
30--50\% was limited to an equivalent total of less than 5 days.  The launch 
vehicle was also maintained at a low humidity in order to prevent condensation 
on ascent \citep{cho94,fri97}.  The high effective area, and particularly the 
good reflectivity down to $\sim1000$  \AA\ \citep{sah00}, are due to both very 
conservative procedures and the extreme vigilance of the engineers and 
scientists assembling the instrument. 

%\begin{figure*}
%\epsscale{0.8} 
%\plotone{fig3.eps}
%\caption{Spectrum of HD93129A in the wavelength interval, 1034.9--1037.6 \AA\,  
%showing the low scattered signal near the center of the interstellar \ion{C}{2} 
%absorption line.  Raw count data integrated over 7371 s from LiF1 are plotted 
%as a function of pixel number in the dispersion direction.   No corrections 
%have been made for the contribution of dark count, air glow or scattered 
%stellar light.}
%\end{figure*}

The \emph{Copernicus} mission showed on-orbit degradation of the Al:LiF 
overcoated optics with time \citep{kal81}.  HUT also showed a slight degradation 
of its SiC coated optics during its second flight \citep{kru99}, which was 
attributed to hydrocarbon contamination in the shuttle environment.  For the 
first three months of the mission, as the \emph{FUSE} instrument outgassed, it 
was pointed at the continuous viewing zone in order to prevent polymerization of 
outgassed hydrocarbons onto the telescope mirrors by ultraviolet light reflected 
from the Earth.  A conservative ram avoidance angle of $>20$\deg\ for the line 
of sight is used to prevent degradation by the small but non-negligible 
abundance of terrestrial atomic oxygen at this altitude.  Although these steps 
slowed observational activity, particularly during the first three months, the 
high effective area and its stability certainly justify the investment; 
measurements of the sensitivity made over a four month period show degradation 
of the effective area to be $\le 5$\%. 

%\begin{figure*}
%\epsscale{1.0} 
%\plotone{fig4.eps}
%\caption{Comparison of \emph{FUSE} spectrum of the QSO HS$1700+64$ at redshift 
%$z=2.743$ obtained in the LiF2 channel with that obtained by HUT. The vertical 
%arrow indicates the onset of the Gunn-Peterson effect \citep{gun65} due to 
%\ion{He}{2} $\lambda$303.78.  Symbols indicate wavelengths where background 
%features have been removed from the data.  Data are binned over 2.5 \AA.}
%\end{figure*}

The profiles of highly absorbed lines are essentially black in the center.
\footnote{Although the K1-16 data in Figure 1 dip almost to zero, they are not 
representative.  The data were obtained as part of an on-orbit scan to map the 
large aperture.  The absorption lines are contaminated by both terrestrial 
airglow and relatively strong nebular emission lines.  Also, when  individual 
spectra were  corrected for wavelength shifts, small artifacts due to the 
adjoining continuum may have been introduced.}
Figure 3 shows a highly absorbed line due primarily to interstellar \ion{C}{2} 
$\lambda$1036.3, \ion{C}{2}$^{*}$ $\lambda$1037.0.  The data have been binned 
over  4 pixels in the dispersion direction (0.027 \AA) and 67 pixels 
perpendicular to the dispersion direction so that the area on the detector 
associated with a data point is $1.6 \times 10^{-4}$ cm$^{2}$.  No corrections 
have been made in the figure for the contribution of dark count, scattered 
airglow or scattered stellar light.  The continuum is sloped due to a \ion{O}{6} 
$\lambda\lambda$1031.9, 1037.6 P Cygni profile and rises from $\sim$ 1400 counts 
at pixel 8150 to about $\sim$ 3500 counts near pixel  8700.  The dashed line in 
the lower panel is drawn at 4.5 counts. The detector dark background 
\citep{sah00} is $\sim 0.8$ counts s$^{-1}$cm$^{-2}$ and hence the contribution 
to the line center is 0.9 counts.  If the remainder is due solely to scattered 
stellar light, it is $<$ 0.2\%\ of the average nearby continuum.

  Although \emph{FUSE} was designed to observe relatively 
bright AGNs and QSOs with fluxes of 1--2 $\times 10^{-14}\ \OIGS$ as sources 
for absorption spectroscopy, the sensitivity is such that it can detect much 
fainter objects. However wavelength binning with an accompanying reduction in 
velocity resolution may be necessary to reduce observation times.  Figure 4 
compares the spectrum  of the QSO HS$1700+64$ obtained by the \emph{FUSE} LiF2 
channel with that from HUT \citep{dav96}.  The HUT instrument was designed for 
spectroscopy of faint objects with a low dispersion that kept the detector dark 
signal to a minimum.  The HUT data were taken at night which reduced the effects 
of both scattered airglow and airglow features.  The \emph{FUSE} spectra shown 
contain both day and night data.  Features detected in a $\sim$ 50 ks blank sky 
exposure, probably terrestrial airglow, have been removed at the positions 
marked in the figure.  The background due to dark count was equivalent to 
roughly twice the signal from the QSO at $\lambda > 1140$ \AA.  A second similar 
spectrum was obtained with the LiF1 channel. The \emph{FUSE} results demonstrate 
that flux levels of order $\le5 \times 10^{-16}$ $\OIGS$ can be detected, 
although careful studies of background effects and detector artifacts will be 
necessary before using the data for detailed quantitative work.

\acknowledgements
We thank the many people who have worked so hard to make the \emph{FUSE} mission 
a success.  We acknowledge the leadership of A. Boggess during the early concept 
studies of the mission.  \emph{FUSE} is an \emph{Origins}  Mission and has been 
funded by NASA's Explorer Program in cooperation with the Canadian Space Agency 
and the Centre National d'Etudes Spatiales of France. The FES was supplied by 
Canada and the gratings by France.  \emph{FUSE} was developed  by the Johns 
Hopkins University in collaboration with the University of California, Berkeley 
and the University of  Colorado and is being operated for NASA by the Johns 
Hopkins University.  Financial support has been provided by NASA contract 
NAS5-32985 to the Johns Hopkins University.  More information on the \emph{FUSE} 
Mission is available at \url{http://fuse.pha.jhu.edu} or at 
\url{http://fusewww.gsfc.nasa.gov/fuse/}.

\clearpage

%% If you want to include your art in the paper, use \plotone.

\figcaption[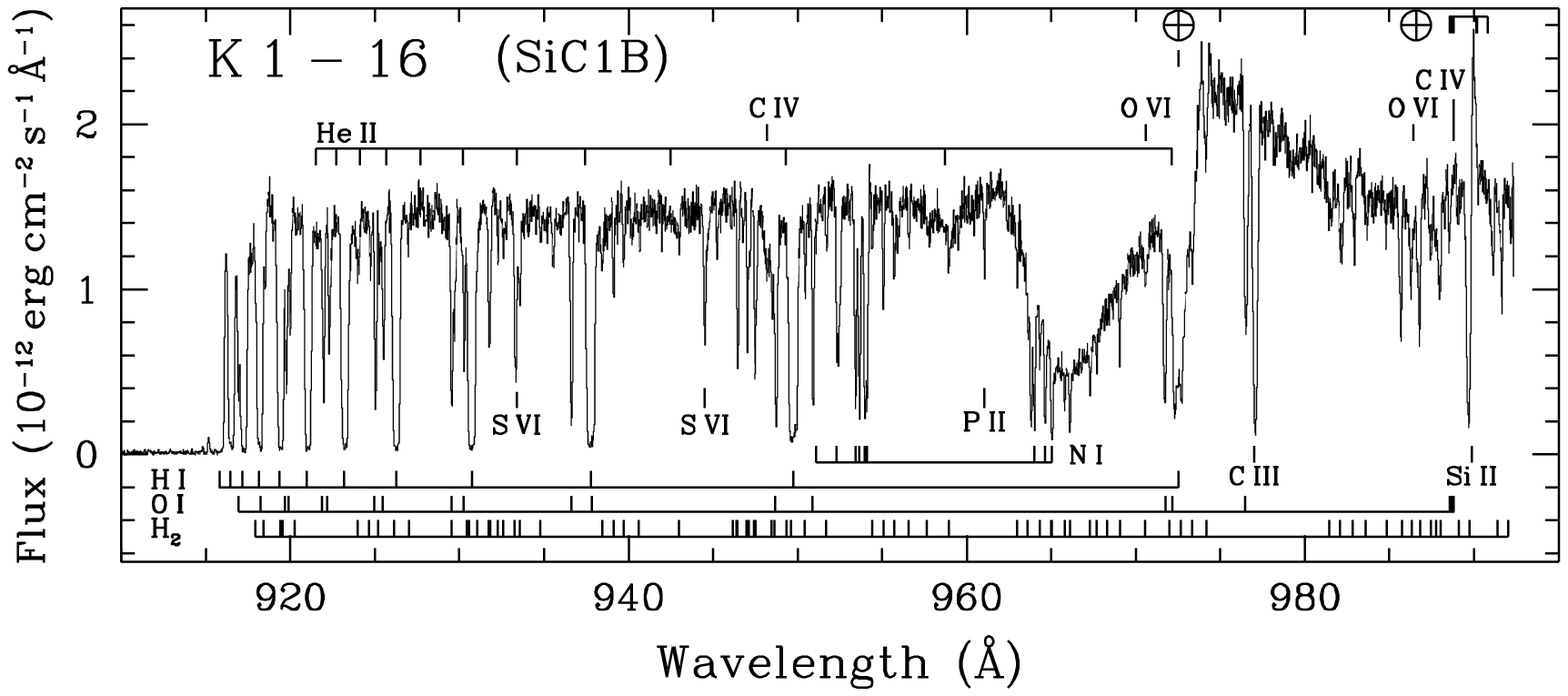 ] {\emph{FUSE} spectrum of the CSPN K1-16, 912--992 \AA.  
Data are from the SiC channel on the detector 1 side.  Transitions in the 
intervening gas  are marked below the spectrum.  Transitions intrinsic to the 
central star are marked above the spectrum. \label{fig1}}

\figcaption[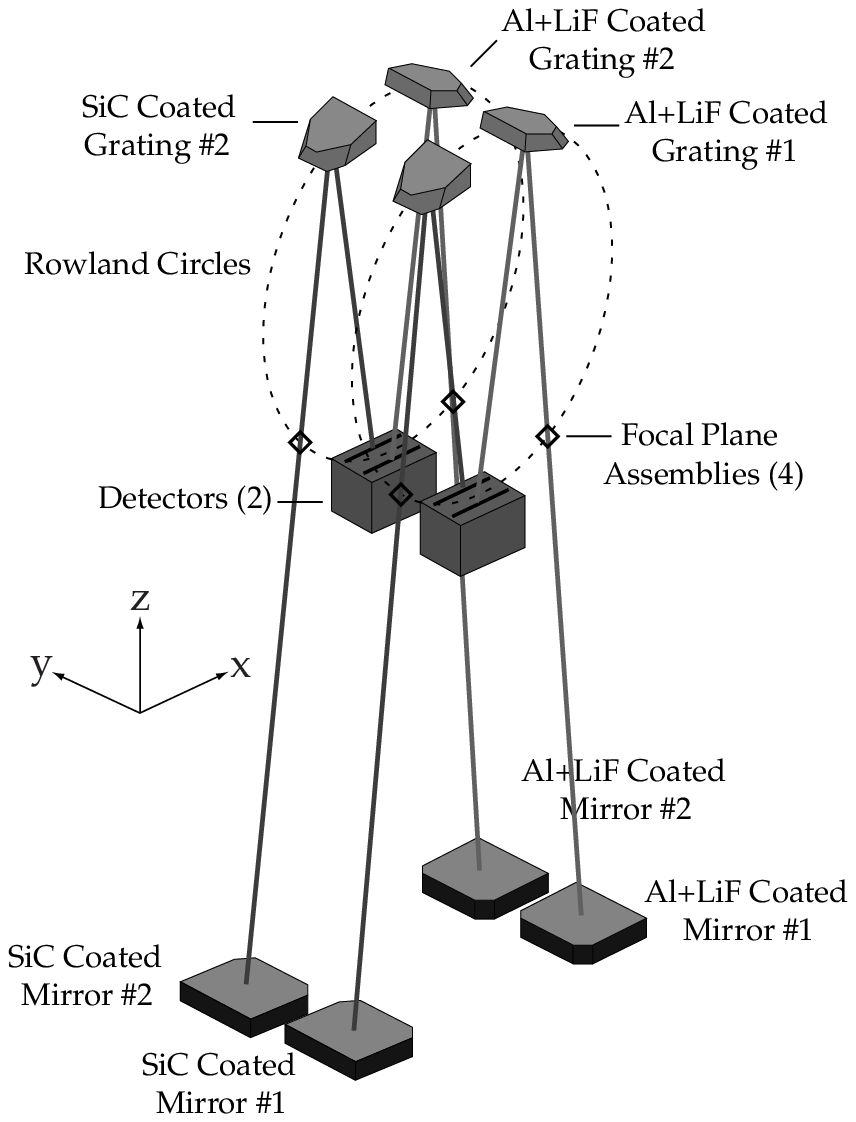] {Schematic of the \emph{FUSE} instrument optical system. 
 The telescope focal lengths are 2245 mm and the Rowland Circle diameters are  
1652 mm.  \label{fig2}}

\figcaption[fig3.eps] {Spectrum of HD93129A in the wavelength interval, 
1034.9--1037.6 \AA\,  showing the low scattered signal near the center of the 
interstellar \ion{C}{2} absorption line.  Raw count data integrated over 7371 s 
from LiF1 are plotted as a function of pixel number in the dispersion direction. 
 No corrections have been made for the contribution of dark count, air glow or 
scattered stellar light. 
\label{fig3}}
 
\figcaption[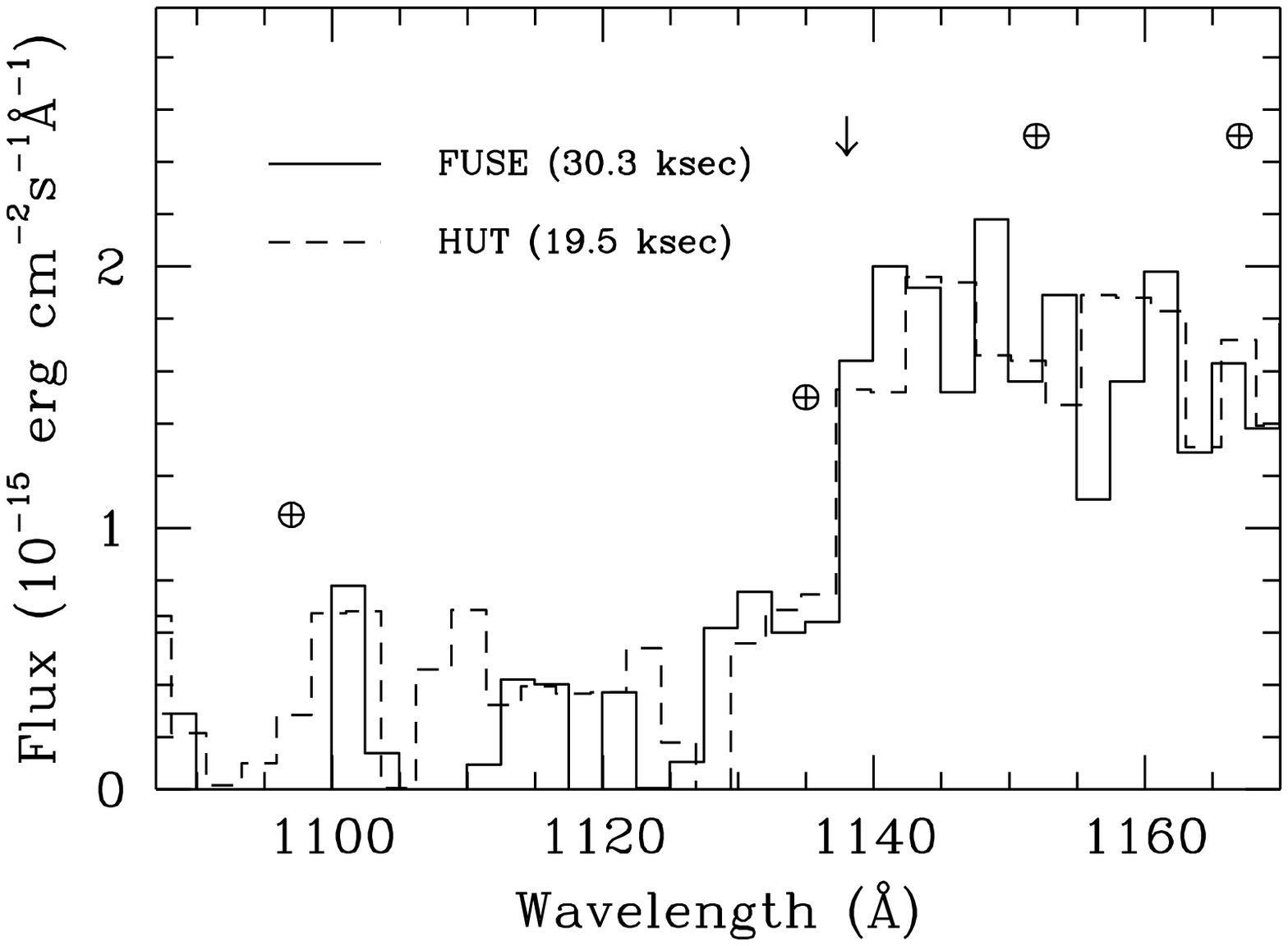] {Comparison of \emph{FUSE} spectrum of the QSO
HS$1700+64$ ($z=2.743$, $V=16.1$) obtained in the LiF2 channel with that 
obtained by HUT.  The vertical arrow indicates the onset of the Gunn-Peterson 
effect \citep{gun65} due to \ion{He}{2} $\lambda$303.78. Symbols indicate 
wavelengths where background features have been removed from the data.  Data are 
binned over 2.5 \AA. \label{fig4}}

\end{document}